\documentclass[3p,times,procedia]{elsarticle}
\usepackage{nupha_ecrc}

\volume{00}

\firstpage{1}

\journalname{Nuclear Physics A}

\runauth{}

\jid{nupha}

\jnltitlelogo{Nuclear Physics A}

\usepackage{amssymb}

\usepackage[figuresright]{rotating}

\newcommand{\nSD}{n_{\textrm{\footnotesize SD}}}
\newcommand{\zcut}{z_{\textrm{\footnotesize cut}}}
\newcommand{\thetacut}{\theta_{\textrm{\footnotesize cut}}}
\newcommand{\qhat}{\hat{q}}

\newcommand{\dif}{\textrm{d}}
\usepackage{hyperref}
\hypersetup{
    colorlinks=true,
    linkcolor=blue,
    filecolor=magenta,      
    urlcolor=cyan,
}
\newcommand{\eqref}[1]{~(\ref{#1})}

\begin{document}

\begin{frontmatter}



\dochead{XXVIIIth International Conference on Ultrarelativistic Nucleus-Nucleus Collisions\\ (Quark Matter 2019)}

\title{Nuclear effects on jet substructure observables at the LHC}


\author[label1]{P. Caucal}
\author[label1]{E. Iancu}
\author[label1]{and G. Soyez}

\address[label1]{Universit\'{e} Paris-Saclay, CNRS, CEA, Institut de physique th\'{e}orique, 91191, Gif-sur-Yvette, France.}

\begin{abstract}
Using a pQCD picture for jet evolution in a dense QCD medium, in which medium-induced parton branchings are factorized from vacuum-like emissions, we study two jet substructure observables: the $z_g$ and the Soft Drop multiplicity distributions. We compute the respective nuclear modification factors using a Monte-Carlo implementation of the parton showers. Our results are in qualitative agreement with LHC data for Pb+Pb collisions. We identify the physical mechanisms explaining our results: incoherent jet energy loss, semi-hard medium-induced emissions and the bias introduced by the steeply falling jet spectrum.
\end{abstract}


\end{frontmatter}


\section{Introduction}
\label{intro}

High-$p_T$ jets are promising probes of the quark-gluon plasma (QGP) created in heavy-ion collisions.
In this context, there is an increasing interest in jet observables dealing with the inner structure of jets \cite{Andrews:2018jcm}. By requiring infrared and collinear safety for these substructure observables, one can hope for controlled calculations in pQCD even in the complex environment of a nucleus-nucleus collision and therefore quantitative comparisons with experiments.

In this proceeding, we explore jet substructure, and in particular the $z_g$ and the Soft-Drop (SD) multiplicity distributions within our new picture of jet fragmentation in a dense QCD medium \cite{Caucal:2018dla, Caucal:2019uvr}. In this picture, vacuum-like and medium-induced emissions (MIEs) are factorized from each other, and separately Markovian which allows for straightforward Monte-Carlo (MC) implementation. The vacuum-like emissions (VLEs) are however modified by the presence of the medium in two important respects: (i) there is a vetoed region for VLEs in phase space and (ii) the angular ordering property can be violated by the first emission outside the medium via colour decoherence.

All the results presented here are obtained using a MC which incorporates these ideas \cite{Caucal:2018ofz, Caucal:2019uvr}. These MC calculations are supported by analytical calculations which will be briefly sketched thereafter. Our results are in qualitative agreement with the measurements done by CMS and ALICE \cite{Sirunyan:2017bsd, Acharya:2019djg}.

\begin{figure}[t]
  \centering
  \includegraphics[page=1,height=5.8cm]{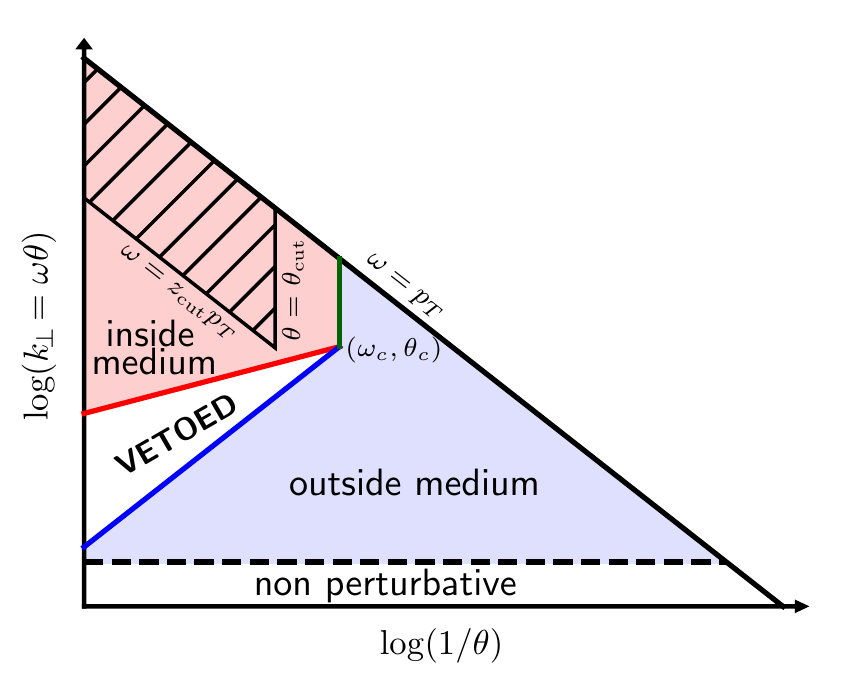}
  \hfill
  \includegraphics[page=2,height=5.8cm]{RAA-v-zg.pdf}
  \caption{\label{raa} \small{Left: Phase space for VLEs in the presence of a dense QCD medium. The ``vetoed'' region is delimitated by the  blue line $t_f=L$ and the red line $k_\perp^2=\qhat t_f$. Right: Our MC results for the jet $R_{AA}$ as a function of $p_{T}$ in bins of $\theta_g$. The dotted line is the inclusive result integrated over all $\theta_g$. The medium parameters $\qhat$ and $L$ are chosen to describe the ATLAS result for the $R_{AA}$ ratio \cite{Aaboud:2018twu}.}
   }

\end{figure}

\section{The $z_g$ and $\nSD$ distributions in the vacuum}

Firstly, we define the $z_g$ and $\nSD$ distributions. 
These observables rely on the SD procedure \cite{Larkoski:2014wba}. For a given jet of radius $R$, SD first reclusters the jet constituents using the Cambridge/Aachen algorithm. The subsequent jet is then iteratively declustered, until the SD stopping condition $z_{12}>\zcut(\Delta R_{12}/R)^{\beta}$ is met,
where $z_{12}$ and $\Delta R_{12}$ are respectively the transverse momentum fraction and the distance in the azimuth-rapidity plane between the two subjets and $\zcut$ and $\beta$ are the SD parameters. One can also impose a lower angular cut-off $\Delta R_{12}>\thetacut$. 
The $z_g$ and $\theta_g$ values are then respectively defined as $z_{12}$ and $\Delta R_{12}$ for the splitting which satisfies the SD condition, if any. The $z_g$-distribution $p(z_g)$ is the cross-section for producing a jet with a given value $z_g$, and we choose to normalize it to the \textit{total} number of jets.

For jets in the vacuum, at double logarithmic accuracy (DLA) with a fixed coupling $\alpha_s$,
\begin{equation}
p(z_g)=\Theta(z_g-\zcut)\int_{\theta_{\textrm{\tiny cut}}}^R\frac{\dif \theta_g}{\theta_g}\frac{2\alpha_s C_R}{\pi}\frac{1}{z_g}e^{-\frac{2\alpha_s C_R}{\pi}A_0(\theta_g)}\textrm{ , with  }A_0(\theta_g)=\int_{\theta{\textrm{\tiny g}}}^R\frac{\dif \theta}{\theta}\int_0^{1/2}\frac{\dif z}{z}\Theta(z-\zcut(\theta/R)^{\beta})
\end{equation}

The number of SD splittings $\nSD$ is found by iterating the SD procedure, following the hardest branch, until the angular cut-off $\thetacut$ is reached \cite{Frye:2017yrw}. $\nSD$ is the number of declusterings passing the SD condition. 
At DLA, $\nSD$ follows a Poisson distribution with average value $2\alpha_sC_R A_0(\thetacut)/\pi$.


\section{Leading medium effects with a ``monochromatic'' jet spectrum}

To simplify, we first consider a leading hard parton with an initial transverse momentum $p_{T0}$ such that $\omega_c< p_{T0}< \omega_c/\zcut$ fragmenting over a distance $L$ into a medium with quenching parameter $\qhat$. The energy scale $\omega_c\equiv\qhat L^2/2$ is the largest energy of a MIE. We expect the medium-jet interactions to modify the $z_g$ and $\nSD$ distributions. Throughout this analysis, we use $\beta=0$, $\zcut=0.1$ and $\thetacut=0.1$. With these choices, SD probes the hatched region in Fig.~\ref{raa}-left, which shows the $(k_\perp\simeq\omega\theta,\theta)$ phase space for VLEs.
\paragraph{Phase space for VLEs} A VLE with formation time $t_f=2/\omega\theta^2$ inside the medium have a phase space bounded by the constraints $k_{\perp}^2=\omega^2\theta^2\ge\qhat t_f$ and $\theta>\theta_c=2/\sqrt{\qhat L^3}$, the coherence angle (see \cite{MehtarTani:2010ma, MehtarTani:2011tz, CasalderreySolana:2011rz}),
whereas a VLE outside must satisfy $t_f>L$. This leads to the existence of a vetoed region in the phase space where no VLE is allowed \cite{Caucal:2018dla}. The vetoed region has an impact on these distributions at DLA e.g. by reducing the area $A_0$ in the parameter of the Poisson law. However, as shown Fig.~\ref{raa}-left, the phase space area probed by SD does not overlap with this vetoed region for our choice of SD parameters. Consequently, the vetoed region can be neglected at DLA.

\begin{figure}[t] 
  \centering
  \includegraphics[page=1,height=5.5cm]{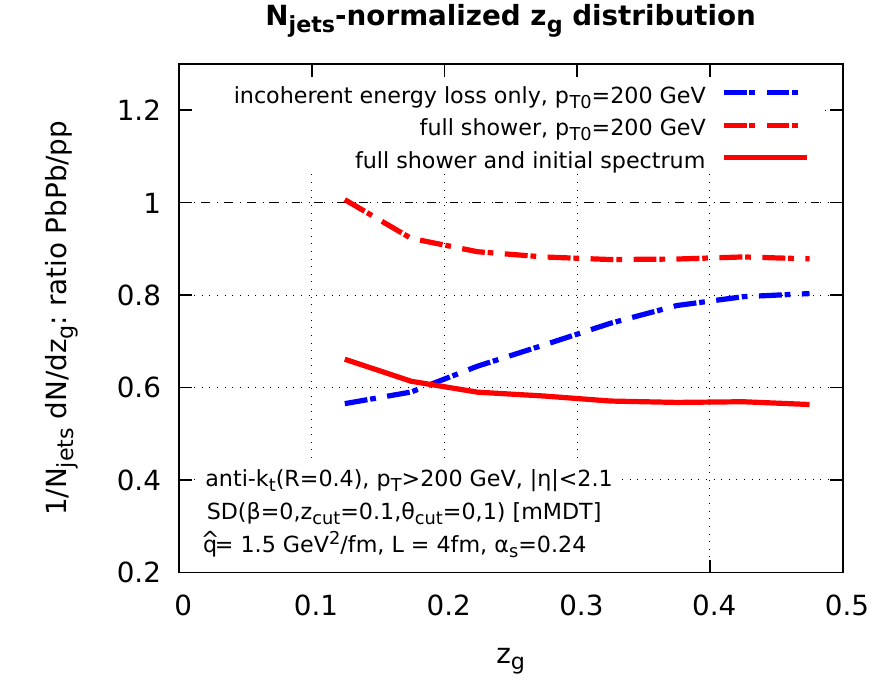}
\hfill
  \includegraphics[page=1,height=5.5cm]{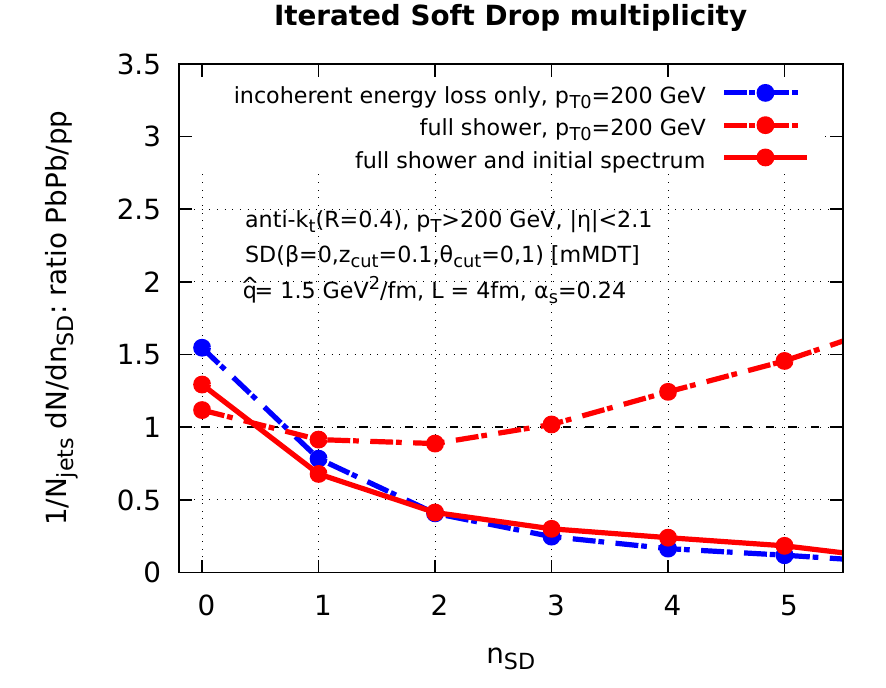}
  \caption{\label{zgnsd} \small{Nuclear modification factor for the $z_g$ (left) and $n_{\textrm{\tiny SD}}$ (right) distributions calculated with our MC. For the dashed curves, the calculation is done for a gluon with $p_{T0}=200$ GeV. For the plain curves, we use the Born-level spectrum in $p_{T0}$. The red curves correspond to the full parton shower described in \cite{Caucal:2019uvr}. For the blue curves, all the MIEs are artificially sent at very large angles to remove the effect coming from intrajet MIEs.}
   }

\end{figure}

\paragraph{Incoherent large angle energy loss}
All the VLEs produced in the in-medium phase space subsequently lose energy via MIEs at large angles before decaying again via brehmsstralung outside the medium. In this picture, all VLEs produced at $\theta\ge\theta_c$ lose energy incoherently.

For a given declustering passing the SD condition (in particular, $\theta_g\ge\thetacut>\theta_c$, see Fig.~\ref{raa}-left), the corresponding $z_g$ value is in general different from the physical energy fraction $z$ before energy loss \cite{Chang:2017gkt}. These two quantities are related via
\begin{equation}
\label{zgvsz}
 z_g\simeq\frac{z p_{T0}-\mathcal{E}_1(z p_{T0},\theta_g)}{p_{T0}-\mathcal{E}_1(z p_{T0},\theta_g)-\mathcal{E}_2((1-z)p_{T0},\theta_g)}
\end{equation}
where $\mathcal{E}_1$ and $\mathcal{E}_2$ are respectively the energy loss of the softer and harder \textit{subjets} via MIEs at angles larger than $\theta_g$. As $\mathcal{E}(p_T)/p_T$ is generally a monotonously decreasing function of $p_T$, Eq.~\eqref{zgvsz} predicts $z_g<z$, with the following consequences for the substructure observables under consideration. The medium/vacuum ratio for the $z_g$-distribution is proportional to $p(z)/p(z_g)\propto z_g/z\simeq 1-\delta z/z_g$ with $\delta z\equiv z-z_g>0$ and this is increasing with $z_g$ ; with our current normalization, it is also smaller than $1$ (blue curve, Fig.~\ref{zgnsd}-left). Furthermore, the logarithmic area probed by SD in phase space is reduced, hence the medium/vacuum ratio for the $\nSD$ distribution decreases with $\nSD$ (blue curve, Fig.~\ref{zgnsd}-right). Yet, these general tendencies can be modified by another medium effect, that we now describe.


\paragraph{Intrajet semi-hard MIEs} 
The semi-hard MIEs with emission angles $\theta<R$ remain inside the jet and can trigger the SD condition \cite{Mehtar-Tani:2016aco}. To estimate the order of magnitude of this effect, we rely on the fact that the spectrum for primary MIEs is well approximated by the BDMPS-Z spectrum \cite{Baier:1996kr, Zakharov:1996fv, Wiedemann:2000za}. Thus, for a jet evolving via primary MIEs only, $\nSD$ is also Poisson-distributed with average value $2\alpha_sC_R A_{m}(\thetacut)/\pi$ and 
\begin{equation}
\label{poissonmie}
 A_{m}(\thetacut)\simeq\sqrt{\frac{\omega_c}{2p_{T0}}}\int_0^{\omega_c/p_{T0}}\frac{\dif z}{z^{3/2}}\int_{\theta_{\textrm{\tiny cut}}}^R\dif\theta\delta(\theta-Q_s/zp_{T0})\Theta(z-\zcut(\theta/R)^{\beta})
\end{equation}
where we approximate the angular distribution with a delta centered around $k_\perp=Q_s\equiv\sqrt{\qhat L}$. When $\beta=0$, this formula enables to distinguish two different regimes \cite{Caucal:2019uvr}. If $\zcut p_{T0}\thetacut\gg Q_s$,  $A_m$ vanishes and substructure observables should not be sensitive to intrajet MIEs. On the contrary, if $\zcut p_{T0}\thetacut\lesssim Q_s$, one finds $A_m\sim(\zcut p_{T0}/\omega_c)^{-1/2}$ which is non-negligible compared to $A_0$ when $\zcut p_{T0}\ll \omega_c$.

In this case, $\nSD$ follows again a Poisson distribution with average value $2\alpha_sC_R(A_{0}+A_{m})/\pi$. Hence the medium/vacuum ratio increases with $\nSD$ (see the red dashed curve in Fig.~\ref{zgnsd}-right). Regarding the $z_g$ distribution, the peak at small $z_g$ seen in our calculation of the nuclear modification of $z_g$ --- the red dashed curve in Fig.~\ref{zgnsd}-left --- is due to the MIEs captured by SD.

\section{Effect of the steeply falling jet spectrum}

Analysing monochromatic jets is helpful to seize the dominant medium effects at play. However, the initial $p_{T0}$ spectrum leads to important modifications of the previous results. 

Due to the steeply-falling underlying $p_{T0}$ spectrum, imposing cuts on the final jet $p_T$ tends to select jets which lose less energy than on the average. That said, 
the more a jet fragments inside the medium, the more it will lose energy at large angles, since the number of sources for MIEs increases \cite{Casalderrey-Solana:2018wrw, Caucal:2019uvr}. Hence, jets with $z_g>\zcut$ or with large $\nSD$ lose more energy than average jets because they have also a larger in-medium multiplicity. Accordingly, they are less likely to be produced in the medium
\cite{Caucal:2019uvr, Casalderrey-Solana2020}. 

Thus, the ratio medium/vacuum for $z_g$ is considerably smaller when using a realistic spectrum, as shown Fig.~\ref{zgnsd}-left,
but the peak at small $z_g$ is still clearly visible. The correlation between $z_g$ and energy loss can be quantitatively measured. To that aim, we highlight that a measurement of the nuclear modification factor for jets $R_{AA}$ for different bins in $\theta_g$ or $z_g$ would be very valuable. We show in Fig.~\ref{raa}-right our Monte-Carlo predictions for such measurement. The important feature of this plot is the striking difference between jets with large $\theta_g$ and small $\theta_g$ with a transition precisely around $\theta_c\simeq 0.04$. 

For the $\nSD$ distribution with $\beta=0$, the effect of the initial cross-section is even stronger, see Fig.~\ref{zgnsd}-right. Jets with large $\nSD$ are highly suppressed and the enhancement seen at large $\nSD$ in the monochromatic case, due to additional MIEs, is no longer visible. Such a compensation implies that one must be cautious when interpreting a measurement of $\nSD$ with $\beta=0$ \cite{Acharya:2019djg}.


\paragraph{Acknowledgements} The work of E.I. and G.S. is supported in part by the Agence Nationale de la Recherche project 
 ANR-16-CE31-0019-01. 
 
\smallskip




\bibliographystyle{elsarticle-num}
\bibliography{refs}







\end{document}